\documentclass{nature}
\usepackage{graphicx}

\title{
\bf 
Vigorous atmospheric motion in the red supergiant star Antares
}

\author{K.~Ohnaka$^{1}$, G.~Weigelt$^2$ \& K.-H.~Hofmann$^2$}
\date{}

\begin{document}

\maketitle

\begin{enumerate}
 \item Instituto de Astronom\'ia, Universidad Cat\'olica del Norte, Avenida
   Angamos 0610, Antofagasta, Chile
 \item Max-Planck-Institut f\"ur Radioastronomie, Auf dem H\"ugel 69, 53121
   Bonn, Germany
\end{enumerate}

{\bf 
Red supergiants represent a late stage of the evolution of stars 
more massive than about nine solar masses, in which they 
develop complex, multi-component atmospheres. 
Bright spots have been detected in the atmosphere of red supergiants using 
interferometric imaging\cite{buscher90,tuthill97,young00,haubois09,baron14}. 
Above the photosphere of a red supergiant, 
the molecular outer atmosphere extends up to about two stellar 
radii\cite{tsuji00a,tsuji00b,perrin04,ohnaka04,harper09,ohnaka09,ohnaka11,ohnaka13,montarges14}. 
Furthermore, 
the hot chromosphere (5,000--8,000 kelvin) and cool gas (less than 3,500 
kelvin) of a red supergiant coexist at about 
three stellar radii\cite{gilliland96,lim98,harper01,harper06}. 
The dynamics of such complex atmospheres has been probed by ultraviolet and 
optical spectroscopy\cite{lobel00,lobel01,gray08,josselin07}. 
The most direct approach, however, is to measure the velocity of gas 
at each position over the image of stars as in observations of the Sun. 
Here we report the mapping of the velocity field over the surface and 
atmosphere of the nearby red supergiant Antares. 
The two-dimensional velocity field map obtained from our near-infrared 
spectro-interferometric imaging reveals vigorous 
upwelling and downdrafting motions of several huge gas clumps at velocities 
ranging from about $-$20 to $+$20 \mbox{km s$^{-1}$}\ in the atmosphere, 
which extends out to about 1.7 stellar radii. 
Convection alone cannot explain the observed turbulent motions and 
atmospheric extension, 
suggesting that an unidentified process is operating in the 
extended atmosphere. 
}

Antares is a well-studied, close red supergiant (RSG) at a 
distance of $170^{+35}_{-25}$~pc (based on the parallax of $5.89\pm 1.00$ 
milliarcsecond = mas, ref. 23).  
We observed Antares 
with the Very Large Telescope Interferometer (VLTI) of the European Southern 
Observatory (ESO) located on Cerro Paranal in Chile 
to directly see the gas motions in the atmosphere. 
The near-infrared VLTI instrument AMBER\cite{petrov07} allowed us to 
record spectrally dispersed interferograms across CO lines between 2.28 and 
2.31~\mbox{$\mu$m}\ with a spectral resolution of 12,000. 
This is sufficient to have approximately 
10 wavelength channels across each CO line and is crucial for directly 
seeing the dynamics of the spatially resolved atmosphere of the star. 
We obtained VLTI/AMBER data 
covering baselines from 4.6 to 82~m and reconstructed
images of Antares at 311 wavelength channels across the observed wavelength 
range (see Methods). 

\begin{figure*}
\begin{center}
\resizebox{\hsize}{!}{\rotatebox{0}{\includegraphics{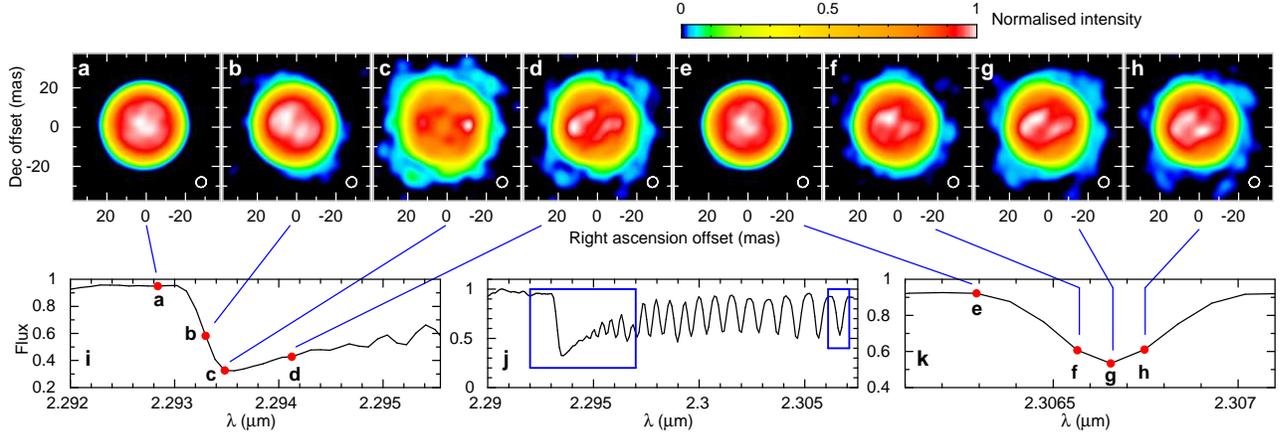}}}
\end{center}
\caption{
Reconstructed velocity-resolved images of Antares. 
The images reconstructed at four wavelength channels near the CO band head 
are shown in panels {\bf a--d}, while those across one of the CO lines are 
shown in panels {\bf e--h}. 
The images in the continuum ({\bf a} and {\bf e}) show the nearly smooth 
stellar disc. 
The images in the CO band head and the CO line ({\bf c}, {\bf d}, {\bf f}, 
{\bf g}, and {\bf h}) reveal two bright spots 
on the stellar disc and the clumpy, extended atmosphere. 
The beam size 
($5.1 \times 5.4$~mas) is shown in the lower right corner of each panel. 
North is up, east is to the left. 
{\bf i--k:} Observed spectrum of Antares. Enlarged views of two wavelength 
ranges (marked with the rectangles in panel {\bf j}) 
are shown in panels {\bf i} and {\bf k}. In these two panels, 
the wavelength channels of the images in panels {\bf a--h} are marked with 
the corresponding alphabetic characters. 
}
\label{alfsco_images}
\end{figure*}

Figure~\ref{alfsco_images} shows the images of Antares reconstructed at 
eight different wavelength channels. 
Thanks to a superb spatial resolution of $5.1\times 5.4$~mas, 
several structures are well resolved with unprecedented spatial and 
velocity accuracy. 
The spatial resolution is about seven times 
finer than the star's angular diameter of $37.61 \pm 0.12$~mas 
in the continuum (see Methods) and nearly 12 times 
finer than the extension of the atmosphere. 
The continuum images 
(Figs.~\ref{alfsco_images}a and \ref{alfsco_images}e) 
show a nearly smooth surface with a weak, large spot at the centre 
with an intensity contrast of 3--4\% (see Methods for the reliability 
of image reconstruction). 
In marked contrast, the images in the CO band head 
(Figs.~\ref{alfsco_images}b, \ref{alfsco_images}c, and \ref{alfsco_images}d) 
as well as those in 
the CO lines (Figs.~\ref{alfsco_images}f, \ref{alfsco_images}g, 
and \ref{alfsco_images}h) clearly reveal two large spots 
with a contrast of $\sim$20\% and an irregularly shaped atmosphere extending 
out to $\sim$1.7 stellar radii ($\approx$32~mas). 
These spots may represent regions with lower CO densities, 
through which the emission from the lower, warmer layers can be seen. 

The previous images of the spotty surface of RSGs 
were taken in the visible, where the strong TiO bands are 
present\cite{buscher90,tuthill97,young00}. 
On the other hand, the image of the well-studied RSG 
Betelgeuse (similar to Antares) 
taken at 833~nm, which better represents the continuum, shows a featureless, 
limb-darkened disc\cite{burns97}.  
The observations at longer wavelengths of 905 and 1290~nm show that 
the spots are weak or absent at these wavelengths, which can be explained by 
smaller TiO opacity\cite{young00} at longer wavelengths. 
The images and modeling of interferometric data of RSGs at 
$\sim$1.6~\mbox{$\mu$m}\ 
show weak to moderate spots presumably due to a jumble of weak 
and moderate lines of CO and CN\cite{haubois09,baron14,montarges16}. 
These results are consistent with our observations in that 
Antares shows a nearly smooth surface in the continuum, and 
the inhomogeneities appear in the CO lines that form in the upper layers 
of the atmosphere. 

From the data cube of the images reconstructed at 311 wavelength channels, 
we extracted the spatially resolved spectrum at each position 
over the surface of the star and the extended atmosphere. 
Figure~\ref{alfsco_specpos} shows the spatially resolved spectra (red lines) 
with a spectral resolution of 8,000 (see Methods) at 
the three representative positions A (on the stellar disc) and B and C 
(in the atmosphere), together with the spatially unresolved spectrum 
(i.e., the spectrum averaged over the entire image; black lines). 
On the one hand, as can be seen in Fig.~\ref{alfsco_specpos}b, 
the spatially resolved spectrum (red line) 
of the bright spot on the surface (position A)
shows stronger absorption lines than the spatially unresolved spectrum. 
On the other hand, the spatially resolved spectra in the 
atmosphere (positions B and C) show the CO lines in prominent emission 
(Figs.~\ref{alfsco_specpos}d and \ref{alfsco_specpos}f)---exactly as 
expected from the Kirchhoff's law. 

A closer look at the spatially resolved line profiles 
reveals that the absorption lines at the position A 
(Fig.~\ref{alfsco_specpos}c, red line) 
is slightly blueshifted with respect to the spatially unresolved spectrum 
(Fig.~\ref{alfsco_specpos}c, black line), 
which means that the gas at this position A is upwelling. 
In Fig.~\ref{alfsco_specpos}e, the peaks of the CO emission lines at the 
position B are clearly 
blueshifted with respect to the absorption lines seen 
in the spatially unresolved spectrum, 
indicating that the gas at the position B is moving towards us. 
In contrast, the CO emission lines at the position C are redshifted with 
respect to the spatially unresolved spectrum (Fig.~\ref{alfsco_specpos}g), 
which means that the gas is moving away from us. 

\begin{figure*}
\begin{center}
\resizebox{16cm}{!}{\rotatebox{0}{\includegraphics{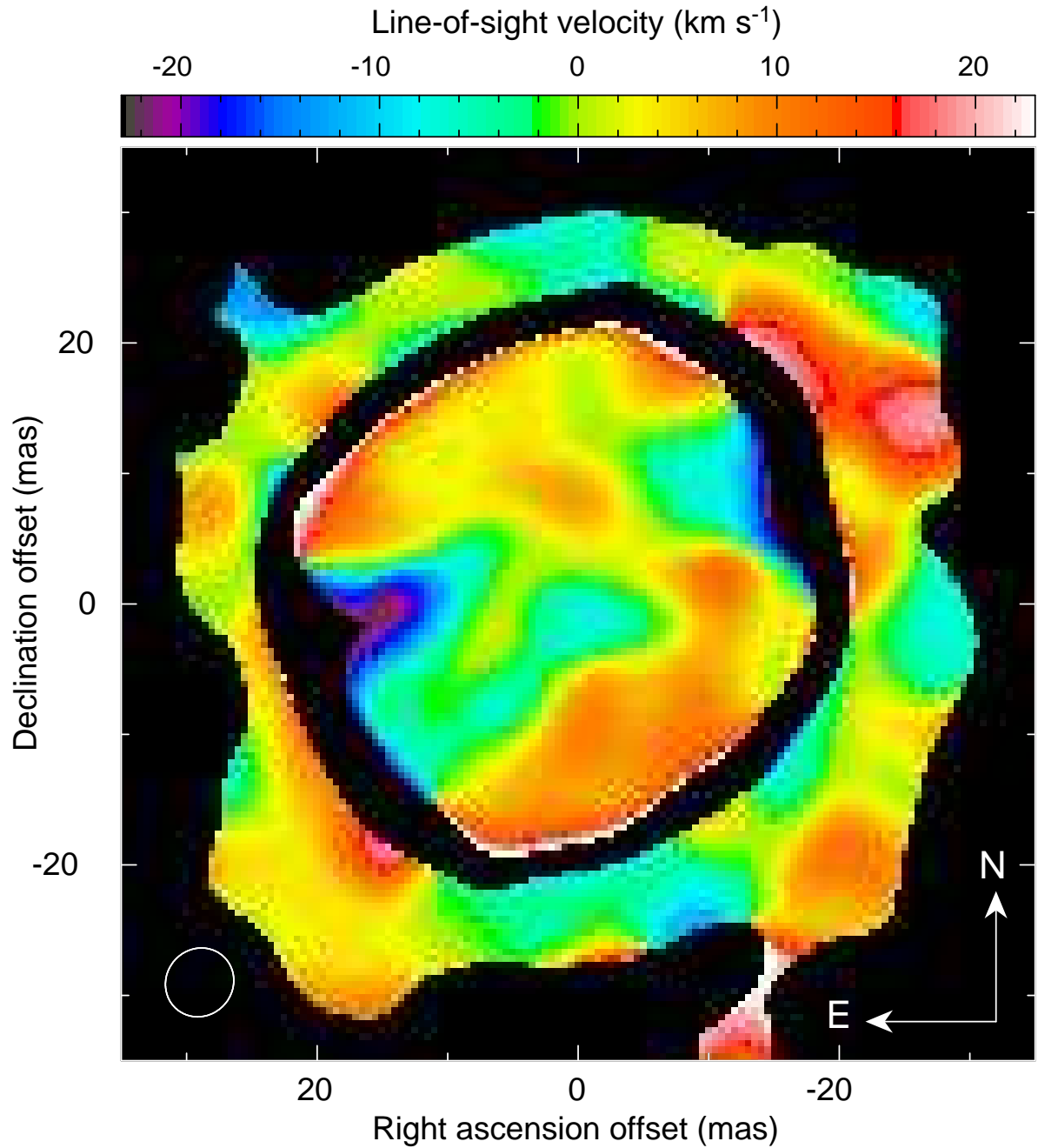}}}
\end{center}
\caption{
Velocity field map over the stellar disc and atmosphere of Antares. 
The velocity is estimated for the positions where the intensity is higher 
than 1.5\% of the peak intensity. 
The black, ring-shaped gap corresponds to the limb of the star, 
where the velocity could not be measured. 
The reason is that the spatially resolved spectra on the limb 
show neither absorption nor emission because the absorption 
and emission cancel out due to the finite beam size (shown in 
the lower left corner). North is up, east is to the left. 
}
\label{alfsco_velmap}
\end{figure*}

We measured the line-of-sight (LOS) gas velocity at each position over 
the stellar disc and atmosphere by calculating the cross-correlation 
between the spatially resolved and unresolved spectra. 
Figure~\ref{alfsco_velmap} shows the two-dimensional velocity field map 
of Antares obtained in this manner (positive and negative velocities 
indicate the gas moving away from us and approaching towards us, 
respectively). 
We resolved 
the velocity field over the stellar disc that reveals upwelling and 
downdrafting at LOS velocities ranging from approximately $-20$ to 
$+10$~\mbox{km s$^{-1}$}\ on a spatial scale of the radius of the star. 
The gas motions in the extended atmosphere are characterised by 
vigorous, inhomogeneous motions of several large clumps at LOS 
velocities ranging from $-10$ to $+20$~\mbox{km s$^{-1}$}. 
We note, however, that 
while the positive and negative LOS velocities observed 
over the stellar disc can be readily associated with downdrafting 
and upwelling motions, respectively, 
there is ambiguity for the clumps observed in the 
extended atmosphere (i.e, outside the limb of the stellar disc). 
For example, the clump in the northwest with an LOS velocity of 20~\mbox{km s$^{-1}$}\ 
may be infalling on the observer's side of the system (i.e., 
in front of the plane perpendicular to the LOS and going through 
the star's centre) or outflowing on the far side of the star (i.e., 
behind the aforementioned plane).

These upwelling and downdrafting motions resemble convection on the 
surface of RSGs\cite{chiavassa11}. 
However, the observationally estimated density in the atmosphere is 
at least six orders of magnitude higher, and the atmospheric extension 
is much larger than predicted by the current convection 
models\cite{ohnaka13,arroyo-torres15}.
This suggests that convection alone cannot lift up the material to 
the observed radius of $\sim$1.7 stellar radii, and 
the CO lines originate in layers higher than the top of convective cells. 
This can explain 
the absence of a correlation between the images in the continuum (probing 
the convection-dominated deep layers) and those in the CO lines (probing 
the extended atmosphere).

Similar upward and downward motions are inferred in the chromosphere of 
Betelgeuse, but the velocity amplitude is much smaller, 
$\sim$5~\mbox{km s$^{-1}$}\ (refs. 19, 20). 
The analysis of optical line profiles in a sample of RSGs 
(but not including Antares) suggests 
upward and downward velocities of up to 17~\mbox{km s$^{-1}$}\ in the upper layers of 
the atmosphere\cite{josselin07}. 
The velocities are comparable to what we measured in Antares, 
although the authors of ref. 22 interpret the motions as originating from 
convection unlike our argument above.  
However, the interpretation of spectral lines obtained by spatially 
unresolved spectroscopy (i.e., averaged over the entire surface and 
atmosphere of a star) in terms of atmospheric motions can be ambiguous. 
For example, a different analysis\cite{gray12} of the radial 
velocities of optical spectral lines of Antares suggests that the convective 
motions penetrate only the lower photosphere (in line with our argument 
above) and that the lines forming in the upper atmosphere show only weak 
atmospheric motions (distinct from our results). 

Since convection alone cannot explain the density and extension of 
the atmosphere, some yet-to-be identified process should be in operation to 
make the atmosphere extended and give rise to the turbulent motions and 
also perhaps the mass loss. 
Given that we did not detect a systematic outflow within 1.7 stellar radii, 
the substantial acceleration of mass loss should take place beyond this 
radius. 
The next challenge remains to identify the driving mechanism 
responsible for the observed turbulent motions. 
Our technique to map out the velocity field over the surface and atmosphere 
of stars other than the Sun 
can be extended to different atomic and molecular lines forming at 
different atmospheric heights. Such tomographic velocity-resolved imaging will 
provide us with a three-dimensional picture of the dynamics of stellar 
atmospheres from deep layers to the outer atmosphere and help us 
identify the process behind the observed atmospheric motions.


\noindent
{\bf Supplementary Information} is linked to the online version of the 
   paper at www.nature.com/nature.

\noindent
{\bf Acknowledgements} 
   We thank the ESO VLTI team for their support for our VLTI/AMBER 
   observations. This work is based on AMBER observations made with the 
   Very Large Telescope Interferometer of the European Southern Observatory 
   (Program ID: 093.D-0468A/B). K.O. acknowledges the grant 
   from the Universidad Cat\'olica del Norte. 

\noindent
{\bf Author contributions} K.O. wrote the telescope proposal and 
   the first paper manuscript, carried out 
   the observations, data reduction, and image reconstruction, and 
   worked on data interpretation. 
   G.W. and \mbox{K.-H.H.} were co-authors on the telescope proposal and 
   worked on data reduction and interpretation. 

\noindent
{\bf Competing Interests} The authors declare that they have no
competing financial interests.

\noindent
{\bf Author Information}  Correspondence and requests for materials
should be addressed to K.O. \\(email: k1.ohnaka@gmail.com).

\clearpage
\renewcommand{\figurename}{Figure}

\clearpage
\begin{figure*}
\begin{center}
\resizebox{\hsize}{!}{\rotatebox{0}{\includegraphics{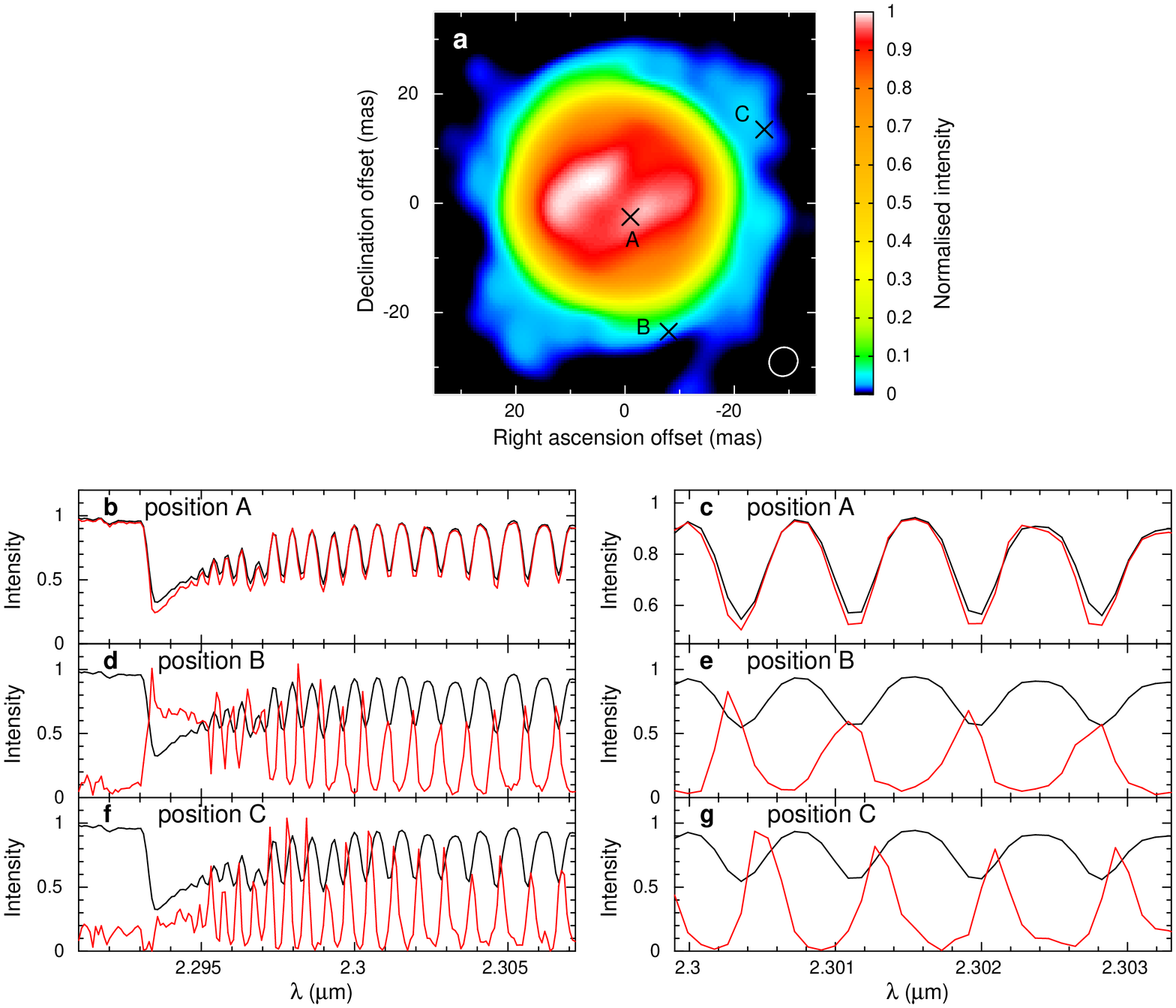}}}
\end{center}
\end{figure*}

\clearpage
\begin{figure*}
\caption{
Spatially resolved spectra over the stellar disc and atmosphere of Antares. 
{\bf a:} Image reconstructed at the centre of the CO line at 
2.30665~\mbox{$\mu$m}\ (north is up, east is to the left). 
The beam size is shown in the lower right corner. 
The crosses mark 
the three positions at which the spatially resolved spectra shown in panels 
{\bf b}--{\bf g} were extracted. 
{\bf b} and {\bf c}: Spatially resolved spectrum obtained at the position A 
on the stellar disc. 
The red lines represent the (scaled) spatially resolved 
spectrum, while the black lines show the spatially unresolved spectrum 
(i.e., spectrum averaged over the entire image). 
Panel {\bf b} shows a comparison for a wide wavelength range, while panel 
{\bf c} shows an enlarged wavelength range across four CO lines. 
{\bf d} and {\bf e}: Spatially resolved spectrum obtained at the position B 
in the atmosphere 
shown in the same manner as in panels {\bf b} and {\bf c}. 
{\bf f} and {\bf g}: Spatially resolved spectrum obtained at the position C 
shown in the same manner as panels {\bf b} and {\bf c}. 
While the spatially resolved spectrum on the stellar disc (A) shows 
stronger absorption than the spatially unresolved spectrum, 
the spectra from the atmosphere (B and C) show the CO lines in prominent 
emission. 
Moreover, 
the emission lines at the position B are blueshifted with respect to the 
absorption lines in the spatially unresolved spectrum, 
while those at the position C are redshifted. 
}
\label{alfsco_specpos}
\end{figure*}

\clearpage

\clearpage
\noindent
{\bf Methods}\\
{\bf Observations.} 
Our VLTI/AMBER observations took place on 2014 April 21, 22, 26, 27, and 28 
(UTC), 
using four different Auxiliary Telescope (AT) 
array configurations---A1-C1-D0, A1-B2-C1, D0-H0-I1, 
and D0-G1-H0, as summarized in Extended Data Tables~\ref{obslog} and 
\ref{obslog} (Program ID: 093.D-0468A/B, P.I.: K.~Ohnaka). 
We used the high spectral resolution mode in the $K$ band (HR\_K) with 
a spectral resolution of 12,000 (but binned down to 8,000 as described 
below) from 2.28 to 2.31~\mbox{$\mu$m}\ to observe the CO first overtones lines 
near the $v = 2 - 0$ band head at 2.294~\mbox{$\mu$m}. 
The extreme high brightness of $K = -4.1$ of Antares 
enabled us to detect 
fringes even on the very long baselines without using the VLTI fringe tracker 
FINITO (Antares is too bright for FINITO to work). 
The use of four different array configurations allowed us to extensively 
sample the object at projected baseline lengths from 4.6 to 82~m. 
Moreover, 
each night we observed Antares throughout the entire night to cover position 
angles as widely as possible. 
The extensive baseline length and position angle coverage resulted in 
a good $uv$ coverage as Extended Data Fig.~\ref{alfsco_uv} shows. 
The data sets taken on 2014 April 28 are of poor quality, and therefore, 
we had to discard most of them except for one data set. 
We observed $\alpha$~Cen~A and $\alpha$~Cen~B as interferometric calibrators 
for the compact configurations (A1-C1-D0 and A1-B2-C1) and extended 
configurations (D0-H0-I1 and D0-G1-H0), respectively. 
We adopted 
the angular diameters of $8.314 \pm 0.016$~mas and $5.856 \pm 0.027$~mas 
for $\alpha$~Cen~A and $\alpha$~Cen~B, respectively\cite{kervella03}. 

We reduced the data with amdlib 3.0.8, which is based on the P2VM 
algorithm\cite{tatulli07,chelli09}. It extracts three interferometric 
observables---visibility amplitude (or simply called visibility), 
closure phase, and differential phase---together with the 
(spatially unresolved) spectrum measured with each telescope. 
Although the original data were taken with a spectral resolution of 12,000, 
it turned out to be necessary to bin all the raw data (spectrally dispersed 
interferograms of the object, sky, dark, and P2VM calibration data) with 
a running box car function to gain S/N\cite{ohnaka09}. 
The data binned to 
a spectral resolution of 8,000 have S/N sufficient for the image 
reconstruction. We discarded the 20\% frames with the poorest fringe S/N 
after checking different fringe selection criteria\cite{ohnaka09}. 
The wavelength calibration was carried out using the telluric lines. 
The uncertainty in the wavelength calibration is 
$1.7\times10^{-5}$~\mbox{$\mu$m}\ (2.2~\mbox{km s$^{-1}$}). 
The observed spectra of Antares were spectroscopically calibrated using 
that of $\alpha$~Cen~A with the method described in our previous 
work\cite{ohnaka13}. 

\noindent
{\bf Limb-darkened disc fitting.}
To estimate the angular size of Antares, 
we fitted the visibilities measured at each wavelength channel 
with a power-law-type limb-darkened disc\cite{hestroffer97}. 
Extended Data Figure~\ref{alfsco_lddfit} shows the derived limb-darkened (LD) 
disc angular 
diameter and the limb-darkening parameter $\alpha$ ($\alpha \rightarrow 0$ 
corresponds to a uniform disc) as a function of wavelength. 
The errors in the LD disc diameter and limb-darkening parameter were 
estimated with the bootstrap technique\cite{efron93}. 
The mean limb-darkened disc diameter and limb-darkening parameter 
over the continuum 
channels are $37.61\pm0.12$~mas and $0.16\pm 0.02$, respectively. 
The fit to the observed visibilities as a function of spatial frequency 
at three representative wavelengths 
(continuum, CO band head, and CO line centre) 
is shown in Extended Data Fig.~\ref{alfsco_lddfit_visfreq}. 
The data in the continuum are fairly fitted with the limb-darkened disc. 
However, the median reduced $\chi^2$ of the fit in the continuum is 11, 
which indicates the presence of inhomogeneities over the stellar disc. 
In the CO band head and CO lines, the limb-darkened disc diameter increases 
up to $\sim$45~mas, and the limb-darkening parameter increases to $\sim$1. 
However, the reduced $\chi^2$ of the fit in the CO band head and the CO lines 
is as large as 30, 
which means that the object appears far more complex than a limb-darkened 
disc.

\noindent
{\bf Image reconstruction.}
We used the MiRA package ver.0.9.9\cite{thiebaut08} for the image 
reconstruction of Antares. 
We implemented the self-calibration imaging technique that takes advantage 
of differential phase measurements in addition to closure phases usually used 
in infrared interferometric image reconstruction\cite{ohnaka11,millour11}. 
In this technique, we first reconstruct the images at all continuum 
wavelength channels using the measured visibility amplitude 
and closure phase alone. On the one hand, the Fourier phase in the continuum 
can be computed from the reconstructed images at the continuum wavelength 
channels. 
On the other hand, differential phases measured with AMBER represent the phase 
in spectral lines with respect to the continuum. Therefore, we can restore 
the Fourier phase in the CO lines using the Fourier phase computed in the 
continuum and the differential phases measured in the CO lines. 
Then the image reconstruction is carried out at all wavelength channels 
using the measured visibility amplitudes and closure phases as well as 
the restored Fourier phases. 
The reconstructed images of Antares were convolved with a Gaussian point 
spread function (PSF) 
with a full width at half maximum of $5.1 \times 5.4$~mas, which 
was derived by fitting the central peak of the dirty beam.

As an initial model, we adopted a power-law-type limb-darkened disc 
with the angular diameter of $37.61$~mas and the limb-darkening parameter 
of 0.16 as derived above. 
We used a Fermi-function-type prior 
$Pr(r) = 1/(\exp((r-r_{\rm p})/\varepsilon_{\rm p})+1)$, which was
successfully used in our previous work\cite{ohnaka11,ohnaka13}. 
This is a function of $r$ (radius in mas), and 
the parameter $r_{\rm p}$ defines the radius where the function 
rapidly decreases to 0. The parameter $\varepsilon_{\rm p}$ defines 
the steepness of the decrease. We set $r_{\rm p}$ and $\varepsilon_{\rm p}$ 
to be 22~mas and 1~mas, respectively. 
Combined with the maximum entropy regularization, 
this allows us to reconstruct the extended atmosphere without causing 
significant artifacts at very large distances from the star. 
We carried out the image reconstruction with different values for 
$r_{\rm p}$ and confirmed that the results are not noticeably affected 
up to $r_{\rm p}$ = 30~mas. 
Extended Data Figures~\ref{reconst_fitBH} and \ref{reconst_fitLine} show 
comparison between the measured interferometric observables 
(visibility amplitude, Fourier phase, and closure phase) and those 
from the images reconstructed at eight wavelengths 
channels shown in Fig.~\ref{alfsco_images}. 

We also carried out image reconstruction tests with simulated interferometric 
data generated from a limb-darkened disc with noise comparable to that of 
the Antares data. 
With the simulated image known, these experiments allow us to verify 
whether or not we can reconstruct the original image with the adopted 
reconstruction parameters and also to estimate the level of artifacts.  
We found out that the original 
limb-darkened disc image can be well reconstructed. The residuals after 
subtracting the original limb-darkened disc from the reconstructed 
image (both images convolved with the aforementioned Gaussian PSF) 
are 1.5\%. This also means that the weak, central spot with the 
intensity contrast of 3--4\% seen in 
the continuum images (Figs.~\ref{alfsco_images}a and \ref{alfsco_images}e) 
is partially due to the reconstruction artifact but may also partially 
show a real structure.  
The deviations of the observed visibilities 
from the limb-darkened disc as mentioned in the previous section also indicate 
the presence of inhomogeneities in the continuum images. 
We note that other structures seen in the CO line 
images are stronger than this weak, central spot seen in the continuum images.

To extract the spatially resolved spectrum at each position over the 
stellar disc and the atmosphere, we normalised the reconstructed image 
(convolved with the Gaussian PSF) 
at each wavelength channel so that the flux integrated over the entire 
reconstructed image is equal to the flux observed at the corresponding 
wavelength with AMBER.

\noindent
{\bf Code availability} 
The AMBER data reduction package is available 
at http://www.jmmc.fr/data\_processing\_\\
amber.htm.
The image reconstruction package MiRA is available at 
http://cral.univ-lyon1.fr/labo/perso/eric.\\
thiebaut/?Software/MiRA. 
We have opted not to make the code for the self-calibration image 
reconstruction and the analysis of the reconstructed images 
available. The reason is that the routines are customized for the present 
analysis and cannot be readily applied to other cases. 

\noindent
{\bf Source data availability} 
The data sets generated and analysed during this study are
available from the corresponding author upon reasonable request.

\clearpage
\renewcommand{\figurename}{Extended Data Figure}
\renewcommand{\tablename}{Extended Data Table}
\setcounter{figure}{0}

\clearpage
\begin{figure*}
\begin{center}
\resizebox{16cm}{!}{\rotatebox{0}{\includegraphics{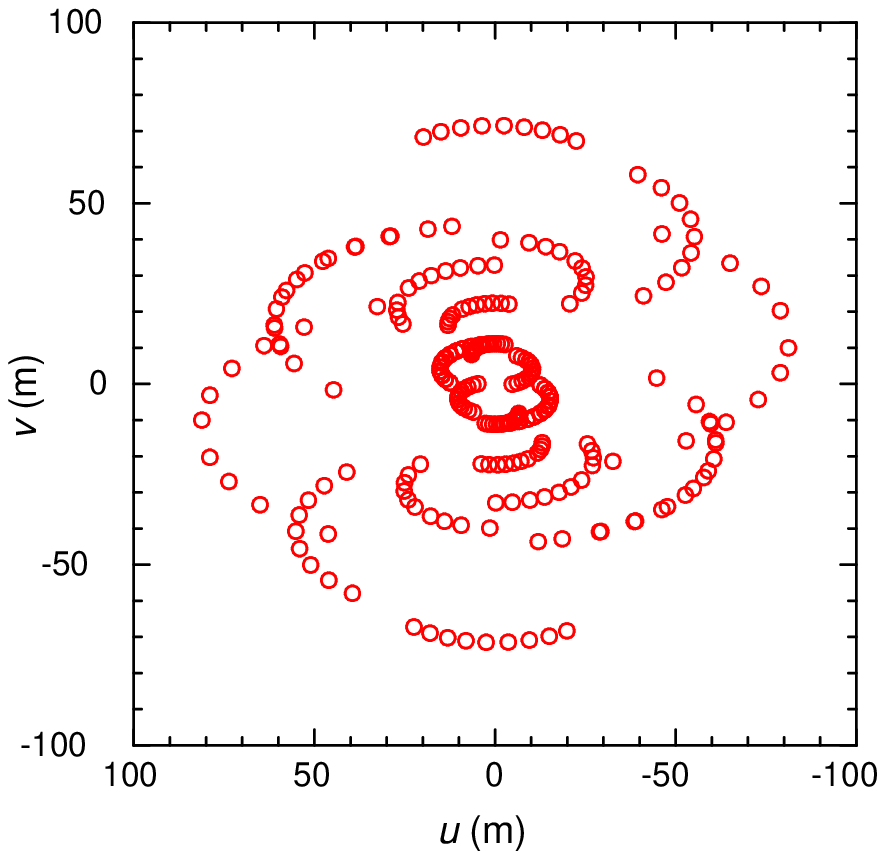}}}
\end{center}
\caption{
The $uv$ coverage of our VLTI/AMBER observations with four different 
AT configurations.
}
\label{alfsco_uv}
\end{figure*}

\clearpage
\begin{figure*}
\begin{center}
\resizebox{\hsize}{!}{\rotatebox{0}{\includegraphics{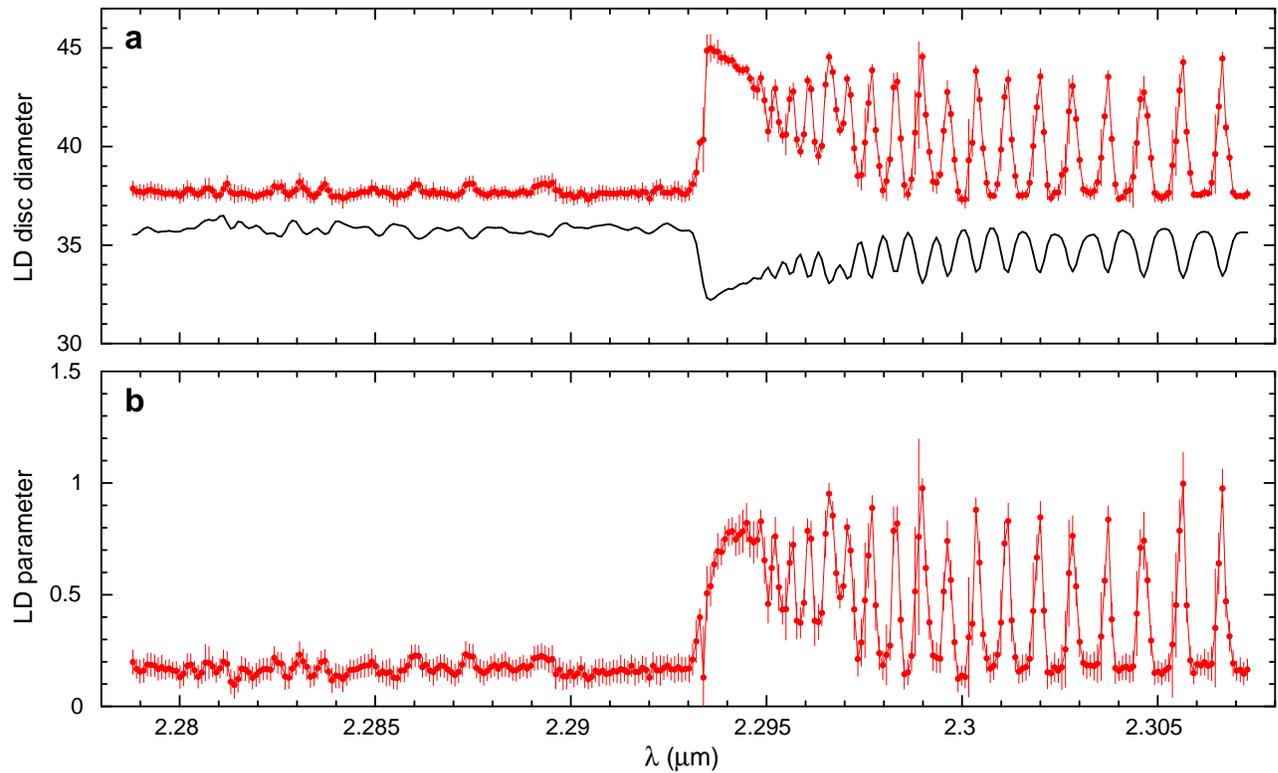}}}
\end{center}
\caption{
Limb-darkened disc fit to the AMBER data of Antares. 
{\bf a:} Power-law-type limb-darkened disc diameter as a function of 
wavelength (red dots). The scaled, observed spectrum is shown in black. 
{\bf b:} Limb-darkening parameter as a function of wavelength. 
In both panels, the error bars represent $1\sigma$. 
}
\label{alfsco_lddfit}
\end{figure*}

\clearpage
\begin{figure*}
\begin{center}
\resizebox{\hsize}{!}{\rotatebox{0}{\includegraphics{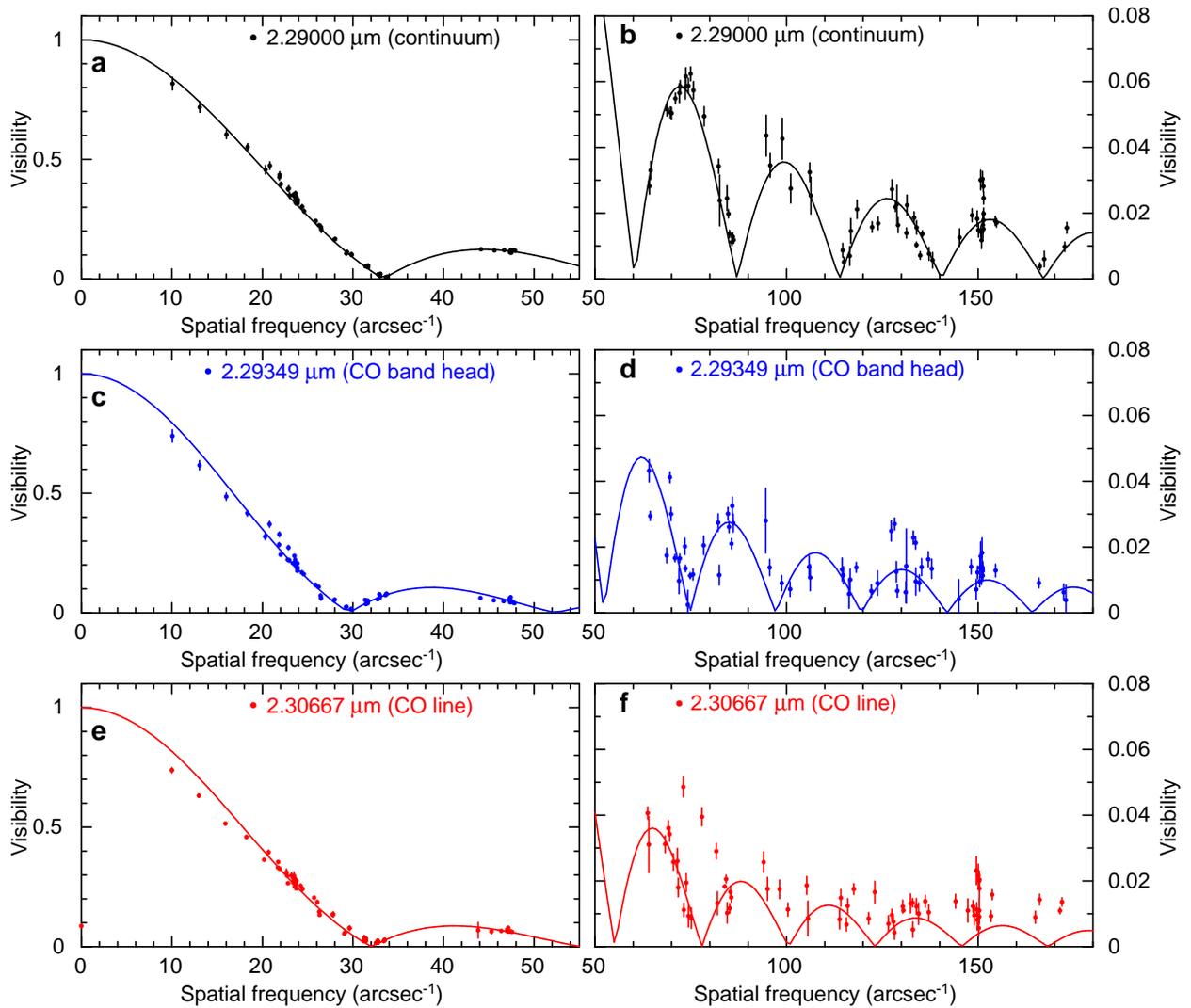}}}
\end{center}
\caption{
Limb-darkened disc fit to the observed visibilities as a function of 
spatial frequency. 
The fit at a wavelength channel in the continuum, in the CO band head, and 
at the center of one of the CO lines is shown in panels 
({\bf a} and {\bf b}), ({\bf c} and {\bf d}), and 
({\bf e} and {\bf f}), respectively. 
The observed visibilities are plotted 
with the dots with the 1$\sigma$ errors computed over 
$N_{\rm f} \times N_{\rm exp}$ frames as listed in Extended Data 
Tables~\ref{obslog} and \ref{obslog}. 
The limb-darkened disc fit is shown with the curves. 
}
\label{alfsco_lddfit_visfreq}
\end{figure*}

\clearpage
\begin{figure*}
\begin{center}
\resizebox{\hsize}{!}{\rotatebox{0}{\includegraphics{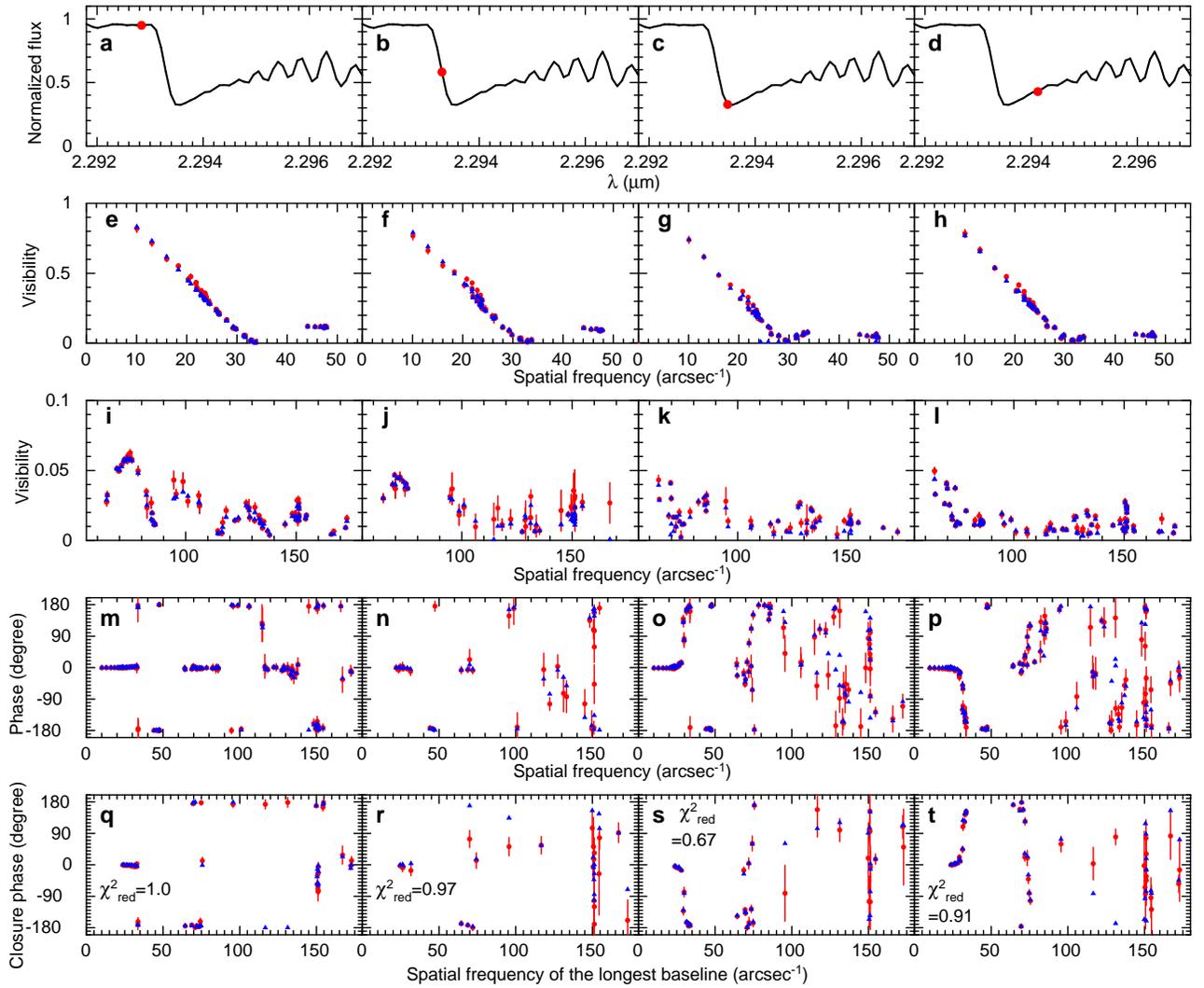}}}
\end{center}
\caption{
Comparison between the measured interferometric observables and those 
from the images reconstructed near the CO band head. 
The filled circles in the top row ({\bf a}--{\bf d}) show the wavelength 
channels in the observed spectrum, which correspond to the wavelengths 
of the images shown in Figs.~\ref{alfsco_images}a--\ref{alfsco_images}d. 
The second, third, fourth, and fifth rows show a comparison for 
the visibility at spatial frequencies lower than 55~arcsec$^{-1}$, 
visibility at spatial frequencies higher than 55~arcsec$^{-1}$, 
Fourier phase, and closure phase, respectively. 
In these panels, the observed data are plotted by the red dots 
with the error bars (1$\sigma$ as described in the legend of 
Fig.~\ref{alfsco_lddfit_visfreq}). 
The blue triangles represent the values from the image reconstruction. 
The reduced $\chi^2_{\rm red}$ values including the visibilities, Fourier 
phases, and closure phases, are given in the panels in the 
bottom row.  
}
\label{reconst_fitBH}
\end{figure*}

\clearpage
\begin{figure*}
\begin{center}
\resizebox{\hsize}{!}{\rotatebox{0}{\includegraphics{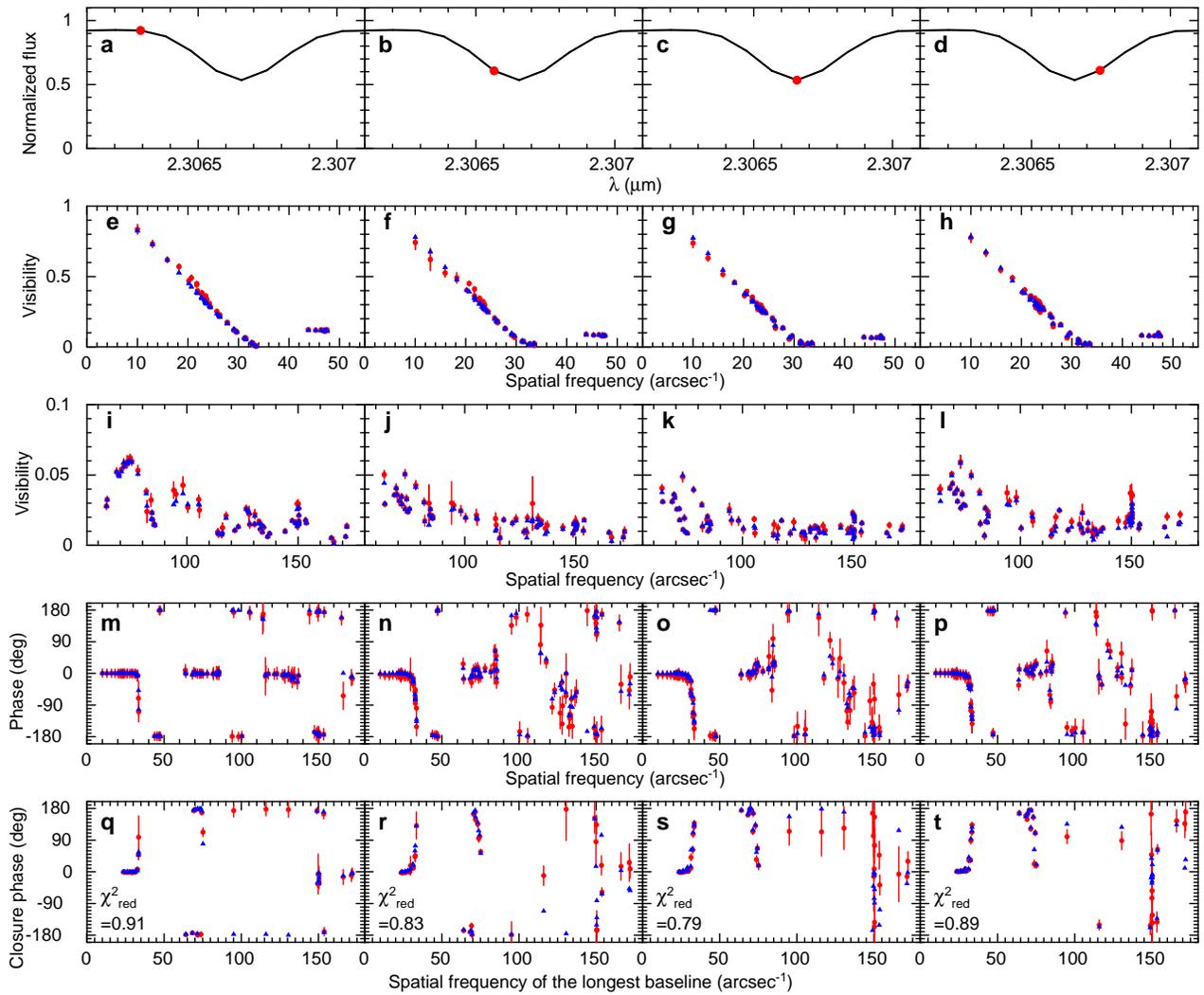}}}
\end{center}
\caption{
Comparison between the measured interferometric observables and those 
from the images reconstructed across one of the CO lines. 
The panels are shown in the same manner as 
Extended Data Fig.~\ref{reconst_fitBH}. 
The filled circles in the top row ({\bf a}--{\bf d}) show the wavelength 
channels in the observed spectrum, which correspond to the wavelengths 
of the images shown in Figs.~\ref{alfsco_images}e--\ref{alfsco_images}h. 
}
\label{reconst_fitLine}
\end{figure*}

\clearpage
\renewcommand{\figurename}{Extended Data Figure}
\renewcommand{\tablename}{Extended Data Table}
\setcounter{figure}{0}

\clearpage
\begin{table*}
\begin{center}
\caption {Summary of VLTI/AMBER observations of Antares. 
$B_{\rm p}$: Projected baseline length. 
PA: Position angle of the baseline vector projected onto the sky. 
$s$: Seeing in the visible. 
$\tau_0$: Coherence time in the visible.  
DIT: Detector Integration Time. 
$N_{\rm f}$: Number of frames in each exposure. 
$N_{\rm exp}$: Number of exposures. 
}
\label{obslog}
\end{center}
\end{table*}

\clearpage
\begin{figure*}
\begin{center}
\resizebox{\hsize}{!}{\rotatebox{0}{\includegraphics{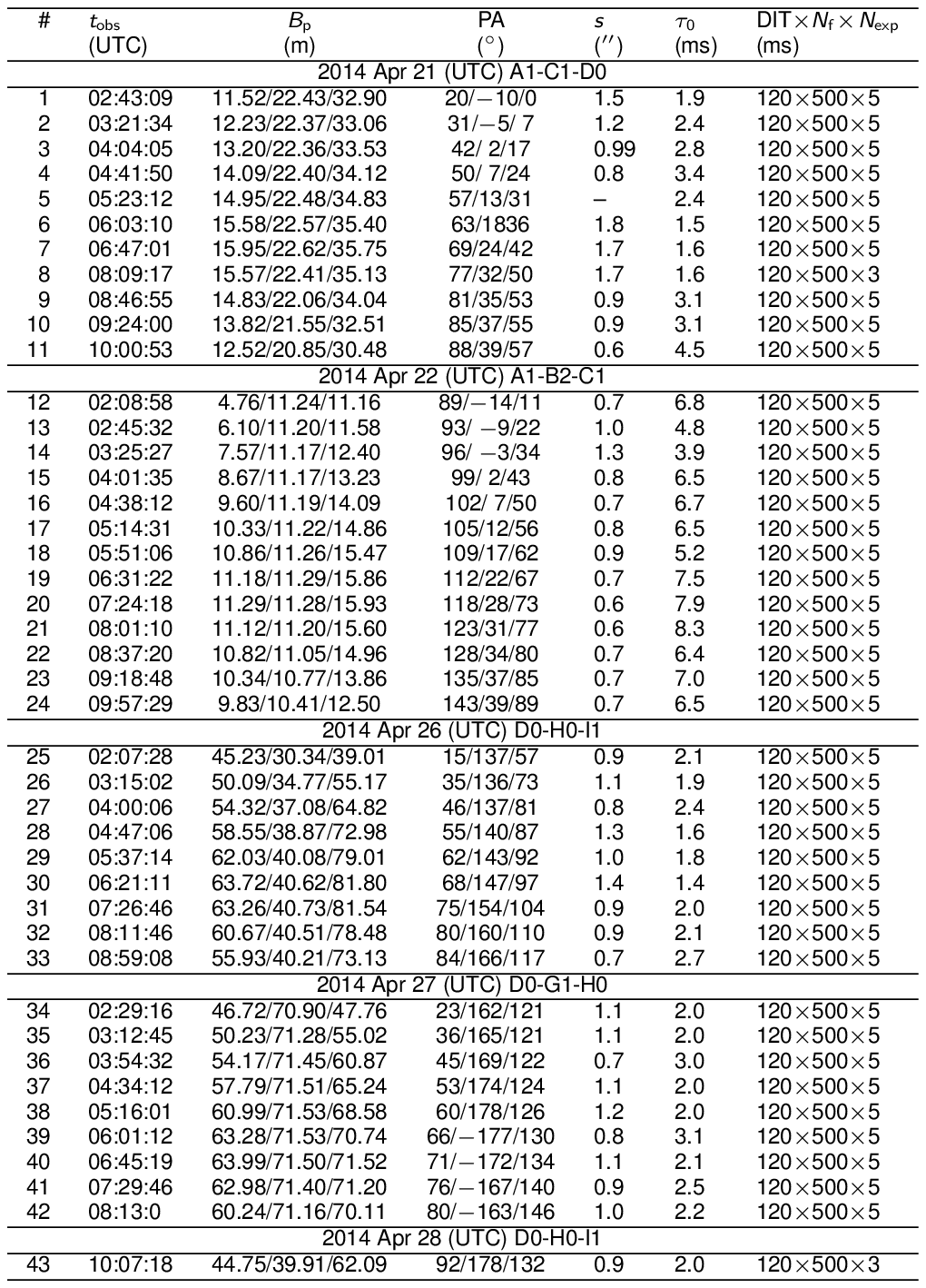}}}
\end{center}
\end{figure*}

\end{document}